\documentclass[aip,jmp,reprint,onecolumn,author-numerical]{revtex4-1}

\usepackage[utf8]{inputenc}
\usepackage[T1]{fontenc}
\usepackage{amsmath,amssymb,calc,amsfonts}
\usepackage{amsbsy}
\usepackage{mathrsfs}
\usepackage{theorem}
\usepackage{graphicx}
\usepackage{color}
\usepackage{wasysym}
\usepackage{latexsym}

{\theorembodyfont{\sffamily} \newtheorem{defi}{Definition:}[section]}

\definecolor{blue}{rgb}{0,0,1}
\definecolor{green}{rgb}{0,0.65,0.5}
\definecolor{red}{rgb}{1,0,0}
\definecolor{vio}{rgb}{0.7,0,1}
\definecolor{ama}{rgb}{1,0.51,0}
\begin{document}

%%%%%%%%%%%%%%%%%%%%%%%%%%%%%%%%%%%%%%%%%%%%%%%%%%%%%%%%%%%%%%%%%%%%%%%%%%%%%%%%
\title{Practical definition of averages of tensors in general relativity}

%%%%%%%%%%%%%%%%%%%%%%%%%%%%%%%%%%%%%%%%%%%%%%%%%%%%%%%%%%%%%%%%%%%%%%%%%%%%%%%%
\author{
Ezequiel F. Boero
and
Osvaldo M. Moreschi
}

\affiliation{
Facultad de Matem\'{a}tica, Astronom\'\i{}a, F\'\i{}sica y Computaci\'{o}n (FaMAF), \\
Universidad Nacional de C\'{o}rdoba, \\
Instituto de F\'\i{}sica Enrique Gaviola (IFEG), CONICET, \\
Ciudad Universitaria,(5000) C\'{o}rdoba, Argentina.
}

%%%%%%%%%%%%%%%%%%%%%%%%%%%%%%%%%%%%%%%%%%%%%%%%%%%%%%%%%%%%%%%%%%%%%%%%%%%%%%%%
\begin{abstract}
We present a definition of tensor fields with are average of tensors over a 
manifold, with a straightforward and natural definition of the
derivative for the averaged fields; 
which in turn makes a suitable and practical construction
for the study of averages of tensor fields that satisfy
differential equations.

Although we have in mind applications to general relativity,
our presentation is applicable to a general $n$-dimensional manifold.
The definition is based on the integration of scalars constructed from a physically 
motivated basis, making use of the least amount of geometrical structure.

We also present definitions of
covariant derivative of the averaged tensors and Lie derivative.
\end{abstract}

\maketitle

%%%%%%%%%%%%%%%%%%%%%%%%%%%%%%%%%%%%%%%%%%%%%%%%%%%%%%%%%%%%%%%%%%%%%%%%%%%%%%%%
\section{Introduction}

It is often the case in the study of complex physical systems to find
information of constituents at a micro scale %%%, 
which influences global phenomena at a macro scale.
In this way, %when 
one is lead to studies of observations that somehow
sample the fields on larger scales, and so one is tempted to use some
kind of averages for its description.

In general relativity the gravitational field itself and the rest of the physical 
fields in a given scale are represented by tensor fields over a curved manifold. 
For example, when large astrophysical or cosmological systems are considered one 
recurs to a macroscopic model for the geometry and the fields suitable for such a 
scale. 
However, the way in which one obtains this description in terms of the 
information of its small constituents it is not in general the consequence
of a clear mathematical procedure.
It is at this point that the need for practical definitions of average of tensors appears.
But the notion of average of tensor quantities in curved spaces 
is a subtle one.

A common feature in any notion of average is associated to the concept of linearity. %; 
More precisely, the basic notion of average involves, in general, the application of 
certain linear operator to the micro quantities in order to obtain the macro ones,
normally by the calculation of some integration.
When one considers averages over a given domain $\Sigma$, often this corresponds
to the integration of some quantity $\mathcal{Q}$, so that
\begin{equation}
\left\langle \mathcal{Q} \right\rangle \propto 
\int_{\Sigma}{\mathcal{Q}\, d\Sigma}.
\end{equation}
This is the case for various fields defined in flat spacetime, for example in electrodynamics
or fluid mechanics. 
In gravitation, on the other hand,
one is confronted with a rich geometrical structure in which various tensors are
related by appropriate derivatives. 
In general, it is not a trivial task to determine which tensors
one should average, and what is the relation among the averaged geometric tensors.
For instance, in the case of the equations for the gravitational field, it is expected that
the curvature tensor associated to an average metric field would differ from the 
average of the curvature tensor of the metric in the micro scale\cite{Ellis:2011hk}.
Also, in the case of the gravitational field equations, one could choose first to average
the energy momentum tensor, that is the curvature, and then integrate for the
corresponding metric. 
This, in general, 
will not coincide with the average of the initial metric either.
However, it seems clear that these difference will be sensible to 
the definition of average that one decides to use. 
Therefore, the possibilities of how one might use the notion of averages are numerous,
and so the definition of average should be concrete, as simple as possible and of practical
use in order to get a clear interpretation of its applications.

Although there are several suggestions for the notion of average of tensors
in the literature\cite{Zalaletdinov:2004wd, Brannlund:2010rs, Behrend:2008az, Green:2010qy},
we were not able to find one that it would meet our expectations to
be simple and of immediate practical physical use.
It is our intention in this work to fill this gap and to provide with a definition
of average of tensors and their corresponding derivatives, 
which could be used in a variety of situations,
and with a minimum requirement of structure.

Before embarking in the details of our presentation of average of tensors in curved spacetimes,
let us motivate our work by mentioning some cases in which the notion of average naturally appears
in the discussion of some important gravitational problems.
They constitute our main guideline for the study of averages in the context of 
general relativity.

\subsection{Motivations}

\subsubsection{Averages in the large scale structure of the spacetime}
The current picture of the Universe includes the
presence of structures in a great range of hierarchical scales which can
be ordered from smaller to bigger; for example:
the solar systems, star clusters, galaxies, galaxy clusters,
supercluster of galaxies, filaments, voids, etc., and finally 
the Universe as whole in its largest scale.
Since gravity is assumed to hold in all the scales of the Universe one is prompted to 
assign a geometry to each characteristic scale. 

In this way one is lead to consider 
the notion of averages in order to connect 
the geometrical description at different scales\cite{Clarkson:2011zq}.
The basic idea being that if one thinks of bigger and bigger volumes, 
the incidence of the large scale distribution of matter on the small scale dynamics
becomes equivalent to the effects that would produce a smoother large scale distribution.

Additionally, one has the information that at early cosmic times matter
was at very high temperature and the distribution of matter at that epoch can be appropriately 
described by a very smooth homogeneous fluid; as is indicated by the CMB observations.

In the largest scale description, which coincides with the
smoother frame, one assumes that the Robertson-Walker geometries
make an appropriate description of the metric of the spacetime.
Most of our knowledge on cosmology comes from the detailed study of this geometry.
However, at the smaller scales one observes highly concentrations of matter
surrounded by empty space. The use of Hilbert-Einstein field equations would 
indicate that for a typical event, at small scales, one would have
a vanishing Ricci tensor and  a non-vanishing Weyl tensor, that is 
\begin{equation*}
 R_{ab} = 0 , \qquad \qquad C_{abc}^{\;\;\;\;d} \neq 0.
\end{equation*}
While in a smooth averaged description on the largest scale, one
would have the contrary; namely,
a non-vanishing Ricci tensor and a vanishing Weyl tensor, that is 
\begin{equation*}
 R_{ab} \neq 0 , \qquad \qquad C_{abc}^{\;\;\;\;d} = 0 ;
\end{equation*}
as is the case in the Robertson-Walker spacetimes\cite{Ellis:2011hk}.

At very small scales, as in the neighborhood of the Sun, it is generally believed
that a typical point of observation will detect zero Ricci curvature.
But, there is no trivial notion of average of tensors that 
will produce a non-zero result from averaging a zero Ricci tensor. 
Then, this situation obviously needs for some detailed study;
in contrast to the fast assumed homogeneous cosmologies 
that are taken for granted as the basis of most cosmological
frameworks.
Having noticed this issue one may wonder whether it is the right idea to
average the energy momentum tensor; since by Hilbert-Einstein equations it determines
the Ricci tensor.

It is often the case in cosmological studies to find implicit use of averages
in the geometrical description of the spacetime metric and curvature.
When we say ``implicit use of average'' we mean that one jumps from a lumpy to 
a completely smooth description without using any mathematical operation.
This issue has been the subject of study for a long time\cite{1984grg..conf..215E}.
Smoothing procedures have been employed
in the study of systems in the framework of 
general relativity, and we should have a good  understanding of all its implications. 

This indicates that one should have precise definitions of the notion of averages
in order to get insight into the way in which one arrives to a coarse grained description 
in general relativity.

\subsubsection{The dark matter problem}\label{subsec:dark}
It is very well known that the use of a Newtonian point of view of the matter
distribution at scales of galaxies confront us with the galaxy rotation curve problem.
This is normally
understood as the need to include some extra non-observed matter,
that would explain the rotational curves of stars around the galaxy.

At the scale of galaxy clusters one also encounters disagreements
among the dynamically inferred matter content, the baryonic expected
matter content and the one estimated by weak gravitational lensing techniques.
Again this problems arise in a Newtonian type description in which
the energy momentum tensor of the matter distribution is completely
described by its timelike component.

Curiously enough, some peculiar geometries\cite{Gallo:2011hi} have been presented
which can explain some aspects of the phenomenology of dark matter in astrophysical systems.
The peculiarity is that in these geometries there appears a non-trivial spacelike
component of the energy momentum tensor.
Since geometrical models, that satisfy the Hilbert-Einstein equations, but which are
not Newtonian in nature, are useful for the description of dark matter phenomena,
one wonders whether the Newtonian point of view is then the correct one for
the description of cosmological systems.
At the same time it also poses the question, how could one understand the extra spacelike component
of the energy momentum tensor.
For example, can the presence of non-negligible space-like component of the energy-momentum tensor 
arise after some averaging process in a big systems composed of many small constituents?

\subsubsection{Averaging the metric}
The metric tensor obviously constitutes one of the candidates to be regarded
in an averaging description, %for the physical description, 
since measurements of lengths and times are determined 
from it, and because it is also essential in the description of the
configuration of the matter fields present in a given scale.

For a big system composed of several subsystems one might consider the idea that
its size and time evolution might be properly determined by a smooth geometry
associated to a coarse grained description of the spacetime. 

This suggests to consider the average of the metric itself from a coarse grain
point of view. In fact, this is what it is normally done at one stroke, in the case of the 
Universe in its largest scale, when one assumes that the metric is homogeneous and isotropic.

More concretely, one can contemplate the case of two typical scales within a multiscale 
hierarchy of the considered system.
A possible approach could be to decompose the metric of the whole system as:
\begin{equation}\label{eq:gbargyh}
g_{ab} = \bar{g}_{ab} + h_{ab};
\end{equation}
with $\bar{g}_{ab}$ a smoother metric of the bigger system and $h_{ab}$ those suitable to the
small scale which could constitute a perturbation to the former.

For example, the application of this idea to a galaxy cluster, that in its bigger
scale it seems to follow more or less an spherical symmetry, will induce one to
take $\bar{g}_{ab}$ from the family of spherically symmetric metrics and use $h_{ab}$ 
to describe the remnant structure.
However, in those cases, normally the procedure does not follow a precise averaging process. 

The fact is that our suggestion above is just one among many different ways in which
one could tackle the issue of averaging the metric. No universal treatment for this
exists, and therefore there are several approaches to the subject.
The common factor in all of them however should be to have available a precise
and practical definition of averaging in order to be able to obtain
concrete conclusions of the calculations.

\subsubsection{Averaging the energy-momentum tensor in a multiscale system}
Instead of considering the averaging of the geometry, one could consider the 
technique of averaging the matter content, previous to the use of the field equation.
The common approach in this respect would be to consider an effective energy momentum 
tensor $\bar T_{ab}$ of the form:
\begin{equation}
\bar{T}_{\alpha\beta} = 
 \frac{1}{V_\Sigma}  
\int_{\Sigma} t_{\alpha\beta} \; d\Omega
;
\end{equation}
where Greek indices denote the components in a given base and $t_{ab}$ is the energy-momentum 
tensor of the small scale constituents that give origin to the big system. 

For example, in a first approach to the cosmological problem one might consider the 
distribution of matter to be determined by the distribution of galaxies.
If matter were only concentrated with visible matter; then at an intermediate point between 
galaxies one would have a zero energy momentum tensor, but its coarse grain average would 
produce a non-zero result. 

The details for a calculation like this would depend on the way one approaches the
problem, and in particular the choice of the region one is integrating.
Normally the characterization of the system would include the choice of a
particular basis in which the calculations have immediate physical interpretation.

\subsubsection{Averaging the connection}
When astrophysical systems are considered, one often encounters very convenient models
that imply the motion of the small bodies on some sort of averaged geometry.
In a galaxy cluster, for example one might obtain information from the dynamics
of each individual galaxy by assuming it is a test body in the background geometry.
Additionally, by making use of the travel of photons across the cluster
one could obtain complementary information from weak lens observations. 
These type of phenomena is described in terms of the geodesic equations;
which actually test the connection, instead of the metric or the curvature.

This suggests that in some cases it might be relevant to consider averages
of the connection over some geometry of reference.
This also would be possible in terms of tensors, since if one has an exact metric $g_{ab}$, 
and some background $\mathbf{g}_{ab}$, then, the difference of their metric connections can be 
described in terms of a tensor, namely, one can write
\begin{equation}\label{eq:conexion}
 \nabla_a v^b = \partial_a v^b + \gamma_{a \;\; c}^{\;\,b} v^c ; 
\end{equation}
where $\nabla_a$ and $\partial_a$ are the metric covariant derivatives
of the metrics $g_{ab}$ and $\mathbf{g}_{ab}$ respectively and 
$\gamma_{a \;\; c}^{\;\,b}$ is a tensor.

Having at hand the definition of derivative of average, one could also 
relate an averaged connection with its corresponding curvature.
Our construction is also suitable for this kind of studies.

\subsection{Our approach}
The previous discussion emphasizes that the issue of averaging tensor fields in
general relativity is rather delicate and necessary. 

In order to have a framework which allows for the consideration of averages one needs to 
have at hand a well defined procedure for performing them and also to know
what are the appropriate quantities to be averaged.
Here we will deal with averages of tensor quantities defined by means of 
integration over spacetime regions.

In this work, we would like to avoid the introduction of extra geometrical structure 
to the plain description of the spacetime or manifold. For this reason, we take a simple
approach and base the notion of averages on the integration of scalars. 
These scalars are defined in terms of a base of vectors which must
be previously specified. 
The choice of this base comes normally from the nature of the problem one is considering. 
For example, in the quasi spherical galaxy cluster mentioned above, one would choose the 
base of vectors using spherical symmetry; which would probably be determined later.

However, it should be stressed that the issue of the best choice of the more appropriate
base is out of the scope of this work. 
For example, if one is interested in the average of a connection in the sense of (\ref{eq:conexion}),
one needs the previous structure that gives the two metrics.
We rather concentrate here only in the mathematical physics aspects of
how one does handle the notion of average of tensor and their derivatives 
in a concrete way, and provide a practical tool for all works 
in which the concept of averaging is needed.

Let us point out that although usually the notion of averages is 
only associated with the choice of a region in which one calculates 
the average. Here we also choose to associate the average to a point; that is we 
require our definition to produce tensor fields. 
All our examples mentioned in our motivations have the
implicit application to a point in the spacetime.
The way in which we introduce this structure is from the assumption that to any 
point of observation $p$ we have at our disposal a specific map to a region of 
integration $\Sigma(p)$, as will be explained in more detail below.
This structure is essential for any attempt to define a derivative of average.

To have available a definition of derivation allows one to treat problems
of averages in which field equations appear.

The article is organized as follows. 
In section \ref{sec:average} we present the definition of average of tensors.
Section \ref{sec:derivative} is dedicated to the notion of derivative of average.
Finally, in the last section \ref{sec:final} we present our final comments and relate our work
with previous approaches of this subject.
In an appendix we develop some algebra of the results appearing in the text.

%%%%%%%%%%%%%%%%%%%%%%%%%%%%%%%%%%%%%%%%%%%%%%%%%%%%%%%%%%%%%%%%%%%%%%%%%%%%%%%%
\section{Definition of averaging of tensor fields}\label{sec:average}

\subsection{The general case}

Consider an $n$-dimensional manifold $\mathcal{M}$ where at each point 
$q \in \mathcal{M}$ a base of vectors $e^{\; a}_{\alpha}$ with its respective 
dual base $\omega_{a}^{\; \alpha}$ is given; that is,
\begin{equation}
e^{\; a}_{\alpha}\,\omega_{\;a}^{\beta} = \delta_{\alpha}^{\;\;\beta};
\end{equation}
where we are using Greek alphabet for numeric base indices, that is
 $\alpha,\beta = 1,...,n$, and we use $a,b,...$ to denote abstract indices.

Therefore, the components of a general tensor field $A^{a...b}_{\;\;\;\;\;\;\;c...d}$
will be denoted in this base as $A^{\alpha...\beta}_{\;\;\;\;\;\;\;\gamma...\delta}$,
such that in this notation one could write
\begin{equation}
A^{a...b}_{\;\;\;\;\;\;\;c...d} = A^{\alpha...\beta}_{\;\;\;\;\;\;\;\gamma...\delta}  
\; e^{\; a}_{\alpha}...e^{\; b}_{\beta} \; \omega_{c}^{\; \gamma}...\omega_{d}^{\; \delta}.
\end{equation}
The components are scalar fields over which the usual average operation in terms 
of integrals can be applied. 

Then, let us choose a reference point $p$, and an assignment of
an $m$-dimensional region of integration which will be denoted as $\Sigma(p)$.
The way in which this assignment is chosen depends on the nature of the problem;
but our construction is independent of the specific assignment.
This assumed map $p \rightarrow \Sigma$ will play a central role in the definition of
derivative of average, presented below.

The definition is:
\begin{defi}\label{def. aver}
{\bf Averages:}
The averaged tensor, denoted by $\left\langle A^{a...b}_{\;\;\;\;\;\;\;c...d} 
\right\rangle_{\Sigma(p)}$, is the new tensor 
\begin{equation}
\left\langle A^{a...b}_{\;\;\;\;\;\;\;c...d} \right\rangle_{\Sigma(p)} = 
\left\langle A^{\alpha...\beta}_{\;\;\;\;\;\;\;\gamma...\delta} 
\right\rangle _{\Sigma(p)} \, e^{a}_{\alpha}...e^{b}_{\beta} 
\, \omega_{c}^{\gamma}...\omega_{d}^{\delta},
\end{equation}
defined at the point $p$; where
\begin{equation}\label{int}
\left\langle A^{\alpha...\beta}_{\;\;\;\;\;\;\;\gamma...\delta} \right\rangle _
{\Sigma(p)} 
\equiv \frac{1}{V_{\Sigma(p)}} \int_{\Sigma(p)}
{ A^{\alpha...\beta}_{\;\;\;\;\;\;\;\gamma...\delta}\, d\Omega_p} ;
\end{equation}
$d\Omega_p $ is the volume element of region $\Sigma(p)$ and $V_{\Sigma(p)}$ is
the volume of such region.
\end{defi}

Some algebraic properties of the definition can be deduced immediately, for instance:
the average of any element of the chosen basis is trivial, namely
\begin{equation}
\left\langle e^{\;a}_{\alpha} \right\rangle_{\Sigma(p)} = e^{\; a}_{\alpha},
\end{equation}
and
\begin{equation}
\left\langle \omega_{a}^{\; \alpha} \right\rangle_{\Sigma(p)} =  \omega_{a}^{\; \alpha} .
\end{equation}
The averages does not commute with the tensor product, that is:
\begin{equation}
\left\langle A_a B^b \right\rangle_{\Sigma(p)} \neq 
\left\langle A_a \right\rangle_{\Sigma(p)}
\left\langle B^b \right\rangle_{\Sigma(p)}.
\end{equation}
%\green{
However notice that for the Kronecker delta tensor one has
\begin{equation}
\left\langle \delta^{\;a}_{b} \right\rangle_{\Sigma(p)} = \delta^{\; a}_{b}
;
\end{equation}
so that the contraction of an averaged tensor
coincides with the average of the contraction, namely
\begin{equation}
\delta^{\;a}_{b}\left\langle A_{a}^{\;b} \right\rangle_{\Sigma(p)} =
\left\langle \delta^{\;a}_{b} A_{a}^{\;b} \right\rangle_{\Sigma(p)}.
\end{equation}

This definition is deliberately constructed in a very simple and straightforward
way to maximize its applicability; having in mind the variety of examples we presented
in the introduction.

\subsection{The case of an $(n-1)$ dimensional region of integration}\label{subsec:n-1-regions}

Taking into account the situations previously considered in the introduction 
it is expected that very often, the regions of integrations might have dimension $n-1$. 
In what follows we would like to present this case. 
We will contemplate the three different cases in which the character of the hypersurface
$\Sigma(p)$ is either timelike, null or spacelike, for applications in general relativity
and Kaluza-Klein theories. 
The case of null hypersurfaces is of particular interest since observable quantities
are determined by the behaviour of the photon on the past null cone.

Let us begin by assuming that a metric and its corresponding form of volume 
$\epsilon_{a_1 \ldots a_n}$ are given in the spacetime; 
then to perform an integration 
over $\Sigma(p)$ we need to work with an $(n-1)$ form of volume induced on such hypersurface, 
namely $\epsilon_{a_1 \ldots a_{n-1}}$. 

In the case in which $\Sigma(p)$ is a null hypersurface,
the choice of $\epsilon_{a_1 \ldots a_{n-1}}$ can be done in the following way.
Within the subspace of 1-forms defined on $\Sigma(p)$ which annihilates all the 
tangent vectors to $\Sigma(p)$ we choose  an element, $\theta_a$ and then define 
$\epsilon_{a_1 \ldots a_{n-1}}$ as
\begin{equation}
\epsilon_{a_1 \ldots a_n} \equiv n \, \theta_{[a_1} \epsilon_{a_2 \ldots a_n]}.
\end{equation}
Let us note that the freedom in this choice is a multiplicative factor of $\theta_a$.
Also, if we denote with $N^a$ a normal to the hypersurface, then
one could take 
\begin{equation}\label{eq:tita}
\theta_a = g_{ab} \, N^b . 
\end{equation}

For example, let us consider the case of the past null cone or future null cone of 
a point in the four dimensional spacetime,
generated by the null vector field $\ell^a$; then, the normal vector $N^a$ is
\begin{equation}
 N^a = \ell^a ,
\end{equation}
and in this notation, one has
\begin{equation}
 \theta_b = \ell_b ,
\end{equation}
when using the null tetrad $(\ell^a, n^a, m^a, \bar m^a)$ one would choose
the three vectors $(\ell^a, m^a, \bar m^a)$  tangent to $\Sigma(p)$.

In the timelike and spacelike cases we have a unique choice of $\theta_a$,
given by eq. (\ref{eq:tita}),
and consequently the volume form $\epsilon_{a_1 \ldots a_{n-1}}$.

If instead of the metric structure one would like to emphasize the coordinate
structure of a coordinate system adapted to the hypersurface $\Sigma(p)$;
one could take, let us say the coordinate $x^1$ to be constant on the hypersurface;
so that the vector $\frac{\partial}{\partial x^1}$ points out of $\Sigma(p)$.
More generally, one could use the notation $\mathscr{N}^a$ for some vector
pointing out of the hypersurface, and then the volume form $\epsilon_{a_1 \ldots a_{n-1}}$ 
will be proportional to the contraction of $\mathscr{N}^a$ with 
$\epsilon_{a_1 \ldots a_{n}}$.
Let us note that we can always choose this vector so that
$\mathscr{N}^a \, \theta_a = 1$.

%%%%%%%%%%%%%%%%%%%%%%%%%%%%%%%%%%%%%%%%%%%%%%%%%%%%%%%%%%%%%%%%%%%%%%%%%%%%%%%%
\section{Derivatives of averages}\label{sec:derivative}
 
As we mentioned in the introduction one of the major goals when an averaging
procedure is specified is to study the differential equations that 
the macro quantities might satisfy. 
An appropriate definition of derivative of averages will
allow us to carry out this task.
In this section we provide a natural definition of derivative 
by applying standard techniques to our definition of the average of a tensor.

In order to take derivatives at a point $p$ it is required the fields 
to be defined in a neighbourhood of $p$; then it is evident that
if a covariant derivative $\nabla_a$ is chosen over the manifold it can be applied 
directly on a field which was previously averaged. 
Therefore, since we need a field in a neighborhood of $p$,
it becomes essential the details of the assignment $p \rightarrow \Sigma(p)$
that determines $\Sigma(p)$ from $p$.
This assignment is arbitrary, but normally fixed by the nature of the problem
in which one is applying the average.

By this assignment from $p$ to a $d$-dimensional (with $d \leqslant n $) 
region $\Sigma(p)$ we mean that there exists a specific map
$\Psi_p$ from a $d$-dimensional set $U$ of $\mathbb{R}^d$ to $\Sigma(p)$.
The details of the map $\Psi_p$ depend normally on the structure available
around the point $p$.
For example, in the case of the four dimensional spacetime, the structure might be given 
by the velocity field $v^a$ of cosmic observers, and their past null cones,
on which any point is referred in terms of the canonical angular coordinates $(\theta,\phi)$,
and affine distance $\lambda$, defined with respect to the velocity field.
Then, to each point $p$ we could assign the region $\Sigma(p)$ defined from the map:
$(\theta,\phi,\lambda \leqslant \lambda^*) \rightarrow$ (points in the past null cone of $p$).

Then, since in the calculation of a derivative, one would compare the value
of an average calculated at a point $p$ with other calculated at nearby point $q$,
one only needs the first order information of how the region $\Sigma(p)$
changes when one moves from $p$ to $q$.
Let us characterize the motion from $p$ to $q$ by a curve $\gamma(s)$,
with tangent vector $\xi^a$. 
In this way one can see that the assignment $p \rightarrow \Sigma(p)$ %,
and the curve $\gamma$ through $p$,
generates curves $\gamma'$ for any other point $p'$ in $\Sigma(p)$.
In other words, the congruence of curves $\gamma'$ is completely determined by the 
generating map $\Psi_p$ mentioned above.
In the example above, the curves $\gamma'$ are generated by maintaining fixed
values of $(\theta,\phi,\lambda)$, and moving from $p$ to $q$.
Therefore, for each assignment $p \rightarrow \Sigma(p)$ and curve $\gamma$
one has a vector field $w^a$, tangent to the $\gamma'$s, with the functional dependence
\begin{equation}
w^a = w^a(p, \xi^a).
\end{equation}
on $\Sigma(p)$.
This field naturally appears in the definitions presented next.

\subsection{Covariant derivative}\label{subsec:cov-derivative}

Given a connection $\nabla_{a}$, we define the covariant derivative in the
direction of the vector $\xi^{a}$ as follows:	
\begin{defi}\label{def. dertens}
{\bf Covariant derivative for the average of tensors:}
\begin{equation}
\begin{split}
\xi^{e}\nabla_{e} \left\langle A^{a...b}_{\;\;\;\;\;\;c...d}\right\rangle_{\Sigma(p)} \equiv &
\,\xi \left( \left\langle A^{\alpha...\beta}_{\;\;\;\;\;\;\gamma...\delta}
\right\rangle_{\Sigma(p)}\right) e^{\;a}_{\alpha}...e^{\;b}_{\beta}\;\omega_{c}^{\;\gamma}...
\omega_{d}^{\;\delta} +  \left\langle A^{\alpha...\beta}_{\;\;\;\;\;\;\gamma...\delta}
\right\rangle_{\Sigma(p)} \xi^{\sigma} \Gamma_{\sigma \; \alpha}^{\;\;\epsilon}\;
e^{\;a}_{\epsilon}...e^{\;b}_{\beta}\;\omega_{c}^{\;\gamma}...\omega_{d}^{\;\delta} \\
& + ...  
+ \left\langle A^{\alpha...\beta}_{\;\;\;\;\;\;\gamma...\delta}
\right\rangle_{\Sigma(p)}\xi^{\sigma}
\Gamma_{\sigma \; \beta}^{\;\;\epsilon}
e^{\;a}_{\alpha}...e^{\;b}_{\epsilon}\;\omega_{c}^{\;\gamma}...
\omega_{d}^{\;\delta} \\
& 
- 
\left\langle A^{\alpha...\beta}_{\;\;\;\;\;\;\gamma...\delta}
\right\rangle_{\Sigma(p)} e^{\;a}_{\alpha}...e^{\;b}_{\beta}\;\xi^{\;\sigma}
\Gamma_{\sigma \;\;\epsilon}^{\;\;\gamma}
\, \omega^{\;\epsilon}_{c}...\omega_{d}^{\;\delta}
 - ...  
 - \left\langle A^{\alpha...\beta}_{\;\;\;\;\;\;\gamma...\delta}
\right\rangle_{\Sigma(p)} e^{\;a}_{\alpha}...e^{\;b}_{\beta}\;
\xi^{\sigma}\Gamma_{\sigma \;\,\epsilon}^{\;\;\delta}\,
\omega_{c}^{\;\gamma}...\omega_{d}^{\;\epsilon};
\end{split}
\end{equation}
where $e^{\;a}_{\alpha}\nabla_{a}e^{\;b}_{\beta}= \Gamma_{\alpha \; \beta}^{\;\;\gamma}e^{\;b}_{\gamma}$
and therefore
$e^{\;a}_{\alpha}\nabla_{a}\omega_{b}^{\;\beta}= -\Gamma_{\alpha \;\; \gamma}^{\,\;\;\beta}
\,\omega^{\;\gamma}_{b}$.
\end{defi}
%%%%%%%%%%%%%%%%%%%%%%%%%%%%%%%%%%%%%

Clearly, this is the most natural notion of derivative applied to the assignment 
$p \rightarrow \Sigma(p)$.

We see that the only term containing the information of how the integration region
$\Sigma(p)$ changes is the first one which has the factor
$\xi \left( \left\langle A^{\alpha...\beta}_{\;\;\;\;\;\;\gamma...\delta}
\right\rangle_{\Sigma(p)}\right)$. 
This derivative can be expressed in terms of the vector field $w^a$ mentioned above. 
In appendix \ref{ap:Aver-Deriv} we use
\begin{equation}\label{lim. deriv}
\xi \left( \left\langle A^{\alpha...\beta}_{\;\;\;\;\;\;\gamma...\delta}
\right\rangle_{\Sigma(p)}\right) =
\lim_{s \to 0} {\frac{1}{s}\left( \left\langle A^{\alpha...\beta}_{\;\;\;\;\;\;\gamma...\delta} 
\right\rangle_{\Sigma(q)} -
\left\langle A^{\alpha...\beta}_{\;\;\;\;\;\;\gamma...\delta} \right\rangle_{\Sigma(p)}  \right)};
\end{equation} 
where $s$ is the parameter along the curve connecting $p$ with $q$ and 
$\xi \equiv \frac{\partial}{\partial s}$, and show that it takes the following 
alternative form: 
\begin{equation}\label{eq:der-second-expression}
\begin{split}
\xi \left( \left\langle A^{\alpha...\beta}_{\;\;\;\;\;\;\gamma...\delta}
\right\rangle_{\Sigma(p)}\right) = \frac{1}{V_{\Sigma(p)}}
\int_{\Sigma(p)}\pounds_{w} \left(  A^{\alpha...\beta}_{\;\;\;\;\;\;\gamma...\delta} 
\, d\Omega_{q(s)} \right) \bigg|_{\|\Sigma(p)} - 
\frac{\left\langle A^{\alpha...\beta}_{\;\;\;\;\;\;\gamma...\delta}
\right\rangle_{\Sigma(p)}}{V_{\Sigma(p)}} \,
\xi\left( V_{\Sigma(q)} \right);
\end{split}
\end{equation}
where $\pounds_{w}$ denotes the Lie derivative along the vector field $w^a$ and
$\|\Sigma(p)$ denotes projection to the tangent space of $\Sigma(p)$. This condition
is needed when the dimensionality of the regions of integrations are lower than $n$.

\subsection{Lie derivative}

Given a vector field $z^a$ we can also consider the Lie derivative of tensors
defined by means of the average \ref{def. aver}. 
%Since tensors so defined have support on the original microscopic manifold
We can apply the standard definition for the Lie derivative 
\cite{Kobayashi63}; which making use of the covariant derivative
can be defined  by: 
\begin{defi}\label{def:dertens}
{\bf Lie derivative for the average of tensors:}

\begin{equation}
\begin{split}
\pounds_{z} \left\langle A^{a...b}_{\;\;\,\quad c...d}\right\rangle_{\Sigma(p)} =& 
z^e \nabla_e \left( \left\langle A^{a...b}_{\;\;\,\quad c...d}\right\rangle_{\Sigma(p)} \right)
-  \left\langle A^{e...b}_{\;\;\,\quad c...d}\right\rangle_{\Sigma(p)} \nabla_e z^a 
- \ldots  
- \left\langle A^{a...e}_{\;\;\,\quad c...d}\right\rangle_{\Sigma(p)} \nabla_e z^b \\
& +  \left\langle A^{a...b}_{\;\;\,\quad e...d}\right\rangle_{\Sigma(p)} \nabla_c z^e 
 + \ldots  
+ \left\langle A^{a...b}_{\;\;\,\quad c...e}\right\rangle_{\Sigma(p)} \nabla_d z^e .
\end{split}
\end{equation}

\end{defi}
 
Let us note that in general, the field $z^a$ will be different from those
used to move the integration region $\Sigma(p)$, even though that both fields
must coincide at $p$.

In the above formula we have not specified the basis one is using in the definition of average;
however, for the particular case in which one considers a coordinate basis, the expression
for the Lie derivative of the average can be given in terms of the numeric index
coordinate basis notation; so that:
\begin{equation}
\begin{split}
\left( \pounds_{z} \left\langle A\right\rangle_{\Sigma(p)} \right)
^{\alpha..\beta}_{\;\;\,\quad \gamma...\delta} &= 
z^\epsilon \partial_\epsilon \left( \left\langle A^{\alpha...\beta}_{\;\;\,\quad \gamma...\delta}\right\rangle_{\Sigma(p)} \right)
-  \left\langle A^{\epsilon...\beta}_{\;\;\,\quad \gamma...\delta}\right\rangle_{\Sigma(p)} \partial_\epsilon 
z^\alpha
- \ldots  
- \left\langle A^{\alpha...\epsilon}_{\;\;\,\quad \gamma...\delta}\right\rangle_{\Sigma(p)} \partial_\epsilon 
z^\beta \\
& +  \left\langle A^{\alpha...\beta}_{\;\;\,\quad \epsilon...\delta}\right\rangle_{\Sigma(p)} \partial_\gamma 
z^\epsilon 
 + \ldots  
+ \left\langle A^{\alpha...\beta}_{\;\;\,\quad \gamma...\epsilon}\right\rangle_{\Sigma(p)} \partial_\delta 
z^\epsilon ;
\end{split}
\end{equation}
where here we use Greek indices to denote the standard numeric coordinate notation.

\subsection{The case of an $(n-1)$ dimensional region}
It is shown in the appendix that for the $(n-1)$ dimensional case, one can prove that for 
a vector field $w$, the derivative of an averaged scalar $A$ can be expressed as:
\begin{equation}
\pounds_{w} \langle A \rangle_{\Sigma(p)}
= \int_{\Sigma(p)} \Bigg[ \nabla_{e}\left( A w^e \right) + 
A\, \theta_f \left(  w^e \nabla_e \mathscr{N}^f 
- \mathscr{N}^e \nabla_e w^f \right) \Bigg] \, d\Omega_p
;
\end{equation}
where we are using the notation introduced above.
This expression can be used in the previous definitions of derivative of averages.

%%%%%%%%%%%%%%%%%%%%%%%%%%%%%%%%%%%%%%%%%%%%%%%%%%%%%%%%%%%%%%%%%%%%%%%%%%%%%%%%

%%%%%%%%%%%%%%%%%%%%%%%%%%%%%%%%%%%%%%%%%%%%%%%%%%%%%%%%%%%%%%%%%%%%%%%%%%%%%%%%
\section{Final comments}\label{sec:final}

In this work we have presented a new definition of average for general tensor fields 
over a manifold which is based on the integration of scalars constructed with the 
use of a preferred basis. 
The techniques used for the definition are designed to be of general application
for a variety of physical interesting situations.
The use of a basis is normally dictated by the nature of the problem one is dealing
with. By using simple available structure, contrary to elaborated 
schemes of other definitions, allows our definition to be of practical and immediate
application. 
Furthermore, our construction also provides with 
a simpler definition of derivative of average\cite{Zalaletdinov:2004wd};
which is essential for the study of physical situations in which 
the averages must be related to differential equations.

Several possibles scenarios have been discussed in our preliminary motivations which
refer to astrophysical and cosmological scales in which the explanation of the
phenomenology might need for techniques involving averages.
In fact, at the largest cosmological scales is where one normally
finds more discussions recurring to the notion of implicit averages.
We think, however, that the appropriate definitions of average of 
tensors might also become very relevant for the study of astrophysical
systems at smaller scales, as for example cluster of galaxies,
individual galaxies, etc.; where dark matter phenomena
is still not very well understood.
It is probably worthwhile to emphasize that for some of these problems
it might become useful to consider the average of the connection,
with respect to some natural background geometry; since several of the observations
come from information of solutions to the geodesic equation, and
it is curious that this technique has not been applied yet.
 
In what follows we would like to relate our definition of average of tensors with 
previous works stressing coincidences and differences. 
For example, in the work of Buchert \cite{Buchert:1999er} which deals with 
inhomogeneous cosmological models where the averages are only applied to meaningful 
scalar quantities present in a $3+1$ decomposition of the Einstein equations and where 
the regions of integration are three dimensional spacelike regions, the author
applies averaging techniques to geometrical equations.
We instead prefer to focus the attention to the notion of average of tensors.
The same author also presents commutation rules for time derivative and average;
we instead present a general definition of derivative of average of tensor.
In a related article  Gasperini et al. \cite{Gasperini:2011us} apply their formalism 
to average scalar quantities over the past light cones, 
making use of a particular choice of observational coordinates in their
suggested physical applications.
This is an instance in which a basis can be extracted for the averages of tensor fields also
and constitute an example of the kind of structure that naturally appears in the study
of specific problems, where the assumed basis in our work are materialized.
However these authors only deal with the simple problem of average of scalars.

Other averaging approaches considering only scalar quantities were presented by Coley 
\cite{Coley:2009yz}, and Kaspar and Sv\'itek \cite{Kaspar:2014cza}. Both are 
essentially concerned with the issue of averages in the cosmological context; 
however they limit their work to the application of averages to
scalars defined from the curvature.

General schemes applicable to any tensor field were proposed by Zalaletdinov 
\cite{Zalaletdinov:1992cf,Zalaletdinov:2004wd, Zalaletdinov:1992cg, Zalaletdinov:2008ts}
which are based on the use of bi-tensors to carry tensors from one point to another within 
the integration region chosen.
The use of bi-tensors can be related to the specification of a basis at each point of the 
manifold. This bi-tensor are required to satisfy certain properties.
In our case, the prescription of a basis on the integration region implies the existence 
of a natural bi-tensor $\mathcal{W}_{a}^{\;b}(p,q)$ as follows 
\begin{equation}
\mathcal{W}_{a}^{\;b'}(p,q) = \omega_{a}^{\; \gamma}(p)\;e_{\gamma}^{\;b'}(q)
;
\end{equation}
which it can be used in the Zalaletdinov approach.

Let us mention also other approaches and definitions used to averages tensor field 
which differ in nature from ours such as those of Green and Wald \cite{Green:2010qy} and 
Behrend \cite{Behrend:2008az}.
The first approach bases their discussion in terms of an existing high frequency 
decomposition of the fields. 

The previous short list does not intend to be a review of all the efforts find
in the literature on works with definitions of derivative of tensor over general
manifolds, instead it is our personal selection to give an idea of the nature
of different approximations to the subject.
Our approach to averages tries to use the minimum structure that allows for 
a practical use of average of tensors.
In this way we provide with a workable definition, along with a definition of derivative of 
average of tensor, that permits to treat general problems in which differential 
equations are involved.
The possible use in studies of the averaged field equations for various physical fields 
and the  application of the averages to astrophysical systems in general and
to cosmological observables will be presented elsewhere.

%%%%%%%%%%%%%%%%%%%%%%%%%%%%%%%%%%%%%%%%%%%%%%%%%%%%%%%%%%%%%%%%%%%%%%%%%%%%%%%%
\subsubsection*{Acknowledgements}

We acknowledge financial support from CONICET and SeCyT-UNC.

%%%%%%%%%%%%%%%%%%%%%%%%%%%%%%%%%%%%%%%%%%%%%%%%%%%%%%%%%%%%%%%%%%%%%%%%%%%%%%%%

\appendix

\section{Formula for the derivative of scalar averaged quantities} \label{ap:Aver-Deriv}

Let us consider the point $p$ which has associated the region of integration 
$\Sigma(p)$ and let us remember that we denote by $d\Omega_p$ the volume element 
of this region.
Also let us consider a curve $\gamma(s)$ connecting $p$ with neighboring
points $q(s)$, each of them having its own region of integration associated, 
namely $\Sigma(q(s))$ with the corresponding volume elements $d\Omega_{q(s)}$.
We will take $\gamma(0) = p$ and $\gamma(s)$ differentiable at $p$ with the 
corresponding tangent vector at $p$ denoted by $\xi^a$.
The smooth change of regions, from $\Sigma(p)$ to $\Sigma(q(s))$ generates 
a flux $\Phi_s : \Sigma(p) \to \Sigma(q(s))$;
which is completely determined by the map $\Psi$, mentioned before, and the
curve $\gamma$.
The vector field $w^a$, is then tangent to this flux, and it
satisfies that at $p$, one has $w^a = \xi^a$.

For the moment,
we will restrict our attention to the cases in which the regions of integrations are 
$(n-1)$-dimensional or $n$-dimensional and we will see that differences arise when 
different dimensions are considered.

Then, our goal is to compute the following limit
\begin{equation}\label{eq:limit-deriv}
\begin{split}
\lim_{s \to 0} \frac{1}{s} &\left[
\frac{1}{V_{\Sigma(q(s))}}\int_{\Sigma(q(s))}A \, d\Omega_{q(s)}
-
\frac{1}{V_{\Sigma(p)}}\int_{\Sigma(p)}A \, d\Omega_p 
\right] \\
&= \lim_{s \to 0} \frac{1}{s}
\left[
\frac{1}{V_{\Sigma(q(s))}}
-
\frac{1}{V_{\Sigma(p)}}
\right]
\int_{\Sigma(p)}A \, d\Omega_p 
+ 
\lim_{s \to 0} \frac{1}{s} \frac{1}{V_{\Sigma(q(s))}}
\left[
\int_{\Sigma(q(s))}A \, d\Omega_{q(s)} 
-
 \int_{\Sigma(p)}A \, d\Omega_p 
\right] \\
&=  - \frac{1}{V_{\Sigma(p)}^2}\lim_{s \to 0} 
\left[
\frac{V_{\Sigma(q(s))} - V_{\Sigma(p)} }{s }
\right]
\int_{\Sigma(p)}A \, d\Omega_p 
+
\frac{1}{V_{\Sigma(p)}} \lim_{s \to 0} \frac{1}{s} 
\left[ 
\int_{\Sigma(q(s))}A \, d\Omega_{q(s)} 
-
\int_{\Sigma(p)}A \, d\Omega_p 
\right]
 ;
 \end{split}
\end{equation}
where the scalar $A$ is any of the scalars coming from the contraction of the tensor 
$A^{a...b}_{\;\;\;\;\;\;c...d}$ with the base 
$\left\lbrace e_{\alpha}^{\; a} , \omega^{\alpha}_{\; a} \right\rbrace$.

The main factor in the second term on the right hand side is:
\begin{equation}
\begin{split}
\lim_{s \to 0} \frac{1}{s} &\left[
\int_{\Sigma(q(s))}A \, d\Omega_{q(s)}
-
\int_{\Sigma(p)}A \, d\Omega_p 
\right] = 
\lim_{s \to 0} \frac{1}{s}
\int_{\Sigma(p)}\left[ 
\Phi^{*}_{s} \left( A \, d\Omega_{q(s)} \right)
-
A \, d\Omega_p 
\right] \\
&= 
\int_{\Sigma(p)} \lim_{s \to 0} \frac{1}{s}
\left[
\Phi^{*}_{s} \left( A \, d\Omega_{q(s)} \right)
-
 A \, d\Omega_p 
\right] 
= \int_{\Sigma(p)} \pounds_{w} \left( A d\Omega_{q(s)} \right) \bigg|_{\|\Sigma(p)}  ;
\end{split} 
\end{equation}	
where $\|\Sigma(p)$ denotes projection to the tangent space of $\Sigma(p)$, 
$\Phi^{*}_{s}$ denotes the pull-back of the flux $\Phi_{s}$,
and $\pounds_{w}$ denotes the Lie derivative along of the vector field $w^a$.

This is the expression that allows us to compute the limit (\ref{eq:limit-deriv}) in 
terms of an integral over $\Sigma(p)$ from which follows eq. (\ref{eq:der-second-expression})
of section (\ref{subsec:cov-derivative}).

%%%%%%%%%%%%%%%%%%%%%%%%%%%%%%%%%%%%%%%%%%%%%%%%%%%%%%%%%%%%%%%%%%%%%%%%%%%%%%%%
\section{The $(n-1)$ dimensional case}\label{ap:n-1}

Following with the algebra, one can see that in the particular case in which the region of 
integration is an $(n-1)$-dimensional submanifold of the $n$-dimensional space, one has from the 
Leibniz rule that	
\begin{equation}
 \pounds_{w} \left( A d\Omega_{q(s)} \right) = 
  \left( \pounds_{w}  A \right) d\Omega_p +  A \pounds_{w} \left(  d\Omega_{q(s)} \right);
\end{equation}
where 
\begin{equation}
\pounds_{w}  A = w^c \nabla_c A,
\end{equation}
and
\begin{equation}
\begin{split}
\pounds_{w} \left(  d\Omega_{q(s)} \right)\bigg|_{\|\Sigma(p)} &= 
\left( \pounds_{w} \epsilon_{b a_2 \ldots a_n} \right) \mathscr{N}^{b} + \epsilon_{b a_2 \ldots a_n} 
\left( \pounds_{w} \mathscr{N}^{b} \right)\bigg|_{\|\Sigma(p)}   \\
&= 
\left( \nabla_e w^e \right) \epsilon_{b a_2 \ldots a_n} \mathscr{N}^{b} + 
\epsilon_{b a_2 \ldots a_n} 
\left( w^e \nabla_e \mathscr{N}^d - \mathscr{N}^e \nabla_e w^d \right)\bigg|_{\|\Sigma(p)}  \\
&=
\left( \nabla_e w^e \right) \epsilon_{b a_2 \ldots a_n} \mathscr{N}^{b} + 
 \theta_f \left(  w^e \nabla_e \mathscr{N}^f - \mathscr{N}^e \nabla_e w^f \right)
 \epsilon_{b a_2 \ldots a_n} \mathscr{N}^d  \\
&= \Bigg[
\left( \nabla_e w^e \right) +  \theta_f \left(  w^e \nabla_e \mathscr{N}^f - \mathscr{N}^e \nabla_e w^f \right) \Bigg] 
d\Omega_p,
\end{split}
\end{equation}
where $\mathscr{N}^a$ is the vector pointing out of $\Sigma(p)$ and $\theta_a$ the 
associated one form mentioned in section (\ref{subsec:n-1-regions})
and satisfying $\mathscr{N}^a \theta_a =1$. 
Let us note that the above formula is valid for any hypersurface regardless of its
nature; that is, for timelike, spacelike or null.

Therefore, we have obtained that
\begin{equation}
\pounds_{w} \left( A d\Omega_p \right) =  \Bigg[ \nabla_{e}\left( A w^e \right) + 
A\, \theta_f \left(  w^e \nabla_e \mathscr{N}^f - \mathscr{N}^e \nabla_e w^f \right) \Bigg] \, d\Omega_p;
\end{equation}
and consequently
\begin{equation}
\pounds_{w} \langle A \rangle_{\Sigma(p)}
=
\lim_{s \to 0} \frac{1}{s}
\left[
\int_{\Sigma(q(s))}A \, d\Omega_{q(s)}
-
\int_{\Sigma(p)}A \, d\Omega_p 
\right] = \int_{\Sigma(p)} \Bigg[ \nabla_{e}\left( A w^e \right) + 
A\, \theta_f \left(  w^e \nabla_e \mathscr{N}^f - \mathscr{N}^e \nabla_e w^f \right) \Bigg] \, d\Omega_p.
\end{equation}

%%%%%%%%%%%%%%%%%%%%%%%%%%%%%%%%%%%%%%%% BIBLIOGRAPHY %%%%%%%%%%%%%%%%%%%%%%%%%%%%%%%%%%%%%%%%

% \bibliography{/home/moreschi/biblio/refosv.bib}
% \bibliographystyle{abbrv}

\end{document}